\documentstyle[aps,prl,epsf]{revtex} 

\begin{document}
\twocolumn[{
\draft

\title{Self-Organized Criticality Induced by Diversity}

\author{ 
\'{A}lvaro Corral, Conrad J. P\'{e}rez, and Albert D\'{\i}az-Guilera 
}
\address{ 
Departament de F\'{\i}sica Fonamental, Facultat de 
F\'{\i}sica, Universitat de Barcelona, Diagonal 647, E-08028 
Barcelona, Spain \\
}
\date{\today}
\maketitle  
\widetext
\begin{abstract}
\leftskip 54.8 pt
\rightskip 54.8 pt
We have studied the collective behavior of a population of 
integrate-and-fire oscillators. 
We show that diversity,
introduced in terms of a random distribution of natural periods, is the
mechanism that permits to observe self-organized criticality (SOC) in the long
time regime.
As diversity increases the system undergoes several transitions from a
supercritical regime to a subcritical one, crossing the SOC region.
Although there are resemblances with percolation,  
we give proofs that criticality takes place for a wide range of values of
the control parameter instead of a single value.
\end{abstract}
\leftskip 54.8 pt

\pacs{PACS numbers: 05.90.+m,87.10.+e,64.60.Ht} 
}]
\narrowtext


In spite of the great interest received along the last decade 
by many systems exhibiting self-organized criticality (SOC)\cite{prl59.381}, 
it is still an open question to
find necessary or sufficient conditions to 
observe this phenomenon, 
since there is no framework to predict, a priori, 
whether an arbitrary extended system will be, by its own dynamics
and without any parameter tuning, critical in the long time regime.  

Nevertheless,
there is a set of common trends which characterize 
systems  displaying SOC \cite{extremal}. 
One of them concerns the dynamics that
drives the elements of the system to a certain threshold. 
When some unit reaches the threshold, 
interaction between elements takes place, 
triggering a chain process or avalanche that ends when all 
the elements 
are below the threshold again. 
Then, a power-law distribution of avalanche sizes is the hallmark of SOC.
A key point, crucial to observe SOC, is the
separation between the slow time scale associated with the process that
leads the units to the threshold (driving), 
and the fast time scale associated with the interaction (avalanches). 
Conservation was also believed essential to obtain SOC. 
Certainly, for the sandpile model \cite{prl59.381}
and other randomly driven models 
it is an indispensable requirement.
A nonconservative dynamics introduces a
characteristic length independent of system size \cite{Manna}. 
However, several continuously
driven models proposed later changed the widespread belief \cite{OFC}. 
In these models SOC is
not necessarily destroyed in a nonconservative regime and the
distribution of avalanche sizes follows a power-law decay in a wide region
of parameter space, with exponents depending on the level of 
dissipation. 

Another point not studied so profoundly concerns the individual features
of each element in the system. Up to now, it has been common to assume
that all the units are identical. 
However, if SOC should have any relevance in
physics or biology it should be robust in spite of the inherent
differences between the members of a population. In other words, diversity
should not destroy the critical properties of a given self-organized 
system. The object of this paper is to show that 
diversity not only does not break SOC 
but as a matter of fact,
it is the mechanism that enables to observe it for certain 
continuously driven models which do not exhibit SOC under normal 
circumstances. 
The subject has a clear general interest. 
There are some collective phenomena such as the mutual
synchronization of the members of a biological population \cite{Buck} which
traditionally have been tackled by assuming that all the units are
identical. However, this assumption is not a necessary requirement.
Several authors \cite{Win,Kura} have shown that after a suitable
modeling, a group of nonidentical oscillators, each endowed with its own
natural frequency, picked from a random distribution,
may display a coherent temporal activity if the
disorder level is below a certain critical value.
A less intuitive opposite behavior has been also reported:
disorder (diversity) can
remove  chaos and foster synchronization in a certain model of oscillators
\cite{Nat}.
Uncorrelated differences between the members of the population trigger
regular spatiotemporal patterns. 
In this paper
we give evidence of another related phenomenon. 
A group of pulse-coupled
oscillators evolve in a complex manner if they are identical, 
generating avalanches with many characteristic
sizes (related with the linear dimension of the system).
However, diversity
will change the collective properties of the long time regime and will
induce SOC. 

Let us consider a population of integrate-and-fire oscillators.
Each
oscillator is defined in terms of a state variable $E$ which evolves in
time as
%
\begin{equation}
\frac{dE_i}{dt} = S - \gamma E_i,
\label{peskin}
\end{equation}
and when $E_{i}$ reaches a threshold value $E_{th}$,
the $i$-th oscillator relaxes, and $E_{i}$
is redistributed instantaneously among its neighbors (labeled by $n$)
according to 
%
\begin{eqnarray}
E_{i} & \rightarrow & 0 \nonumber \\
E_{n} & \rightarrow & E_{n} + \varepsilon_{in},
\label{ffrule}
\end{eqnarray}
and so on for every $E_k \ge E_{th}\ \forall k$.  
This process, that continues until $E_k < E_{th} \ \forall k$,
constitutes an avalanche whose size $s$ is given by the number of 
relaxations (\ref{ffrule}).
If $\sum_n \varepsilon_{in} \le E_{th} \ \forall i$ 
except for at least one element for which the inequality is strict,
it is guaranteed that avalanches of infinite size are impossible.
Since the
relaxing site is reset to zero and a fixed quantity $\varepsilon_{in}$
is transferred to the neighbors, the model is intrinsically nonconservative. 
It is assumed $ \gamma E_{th} < S$ and
$\gamma$ is a non-negative constant whose physical meaning depends on the
model one is dealing with. For instance, this model mimics a simplified
version of the dynamics of spiking neurons, idealizing the cell membrane as 
an $RC$ circuit \cite{Lap}. 
$E_i$ denotes the membrane potential of a given neuron,
$\gamma^{-1}=RC$ the membrane time constant, 
$S$ (in appropriate units) a constant current 
that does act as a driving,
and $\varepsilon_{in}$ the synaptic
coupling strength between neurons $i$ and $n$.
Equations (\ref{peskin}) and (\ref{ffrule}) may
also model the evolution of the cardiac pacemaker \cite{Peskin}, 
swarms of flashing fireflies,
and many
other biological systems \cite{Stro,prl2}.

In order to study local connectivity we have considered the 
case of a two-dimensional square lattice of linear size $L$ 
with nearest-neighbor uniform interactions, 
$\varepsilon_{in} \equiv \varepsilon$. 
Recent studies on integrate-and-fire neurons \cite{Hopherz}
are more devoted to lattice models with periodic boundary conditions.
However, the assumption of open boundary conditions breaks
the homogeneous connectivity allowing the boundary units 
to be connected with less neighbors than the bulk units. 
This assumption will be present in the rest of the paper and
makes the model also interesting in other
fields. 
For $\gamma=0$ it reduces to a coupled
map lattice proposed by Feder and Feder as a stick-slip
model of earthquakes \cite{FF}. 

In addition, we have introduced diversity in terms of a random
distribution of intrinsic periods. 
The period of each oscillator is
given by 
\begin{equation}
    T = \frac{1}{\gamma} \ln \left(\frac{S}{S-\gamma E_{th}}\right).
\end{equation}
There are different ways to
introduce such type of quenched disorder in the model. One possibility is
to assume a distribution of $R$ and $C$. 
Another option is to consider a distribution of input currents $S$. 
Both situations are plausible from a realistic point of view, 
but we have considered the latter.
Let us mention that the most usual way to introduce diversity 
in this sort of models is by assuming a quenched random distribution of
thresholds \cite{Jan,Tor}.
Although the distribution of thresholds
also implies diversity in the intrinsic periods, 
it has influence in both, the slow and the fast time scale, while
our approach only affects the slow dynamics. 
This difference is crucial as we will see later.

When all the oscillators are identical the model described by (\ref{peskin})
and (\ref{ffrule}) does not display SOC for any value of $\gamma \geq 0$ and
$ 0 < \varepsilon \le 0.25$. 
Starting with uniformly distributed random initial conditions,
$E_i \in [0, E_{th} \equiv 1]$,
for $\gamma=0$ and integer ratio $E_{th}/\varepsilon$ only large
avalanches take place because many units reach the threshold
simultaneously \cite{tesikim,review}.
If the ratio $E_{th}/\varepsilon$ is not
an integer, avalanches of all sizes are observed, but they are not
power-law distributed \cite{review}. 
For $\gamma > 0$ (convex driving)
the model exhibits a complex behavior which, depending on the particular 
values of $\gamma$ and $\varepsilon$ ranges from synchronization 
(in the sense that all the avalanches are exactly of size $L^2$) 
to events of all sizes 
distributed in a complicated way, as Fig. 1 illustrates. 
Here we observe that the probability density $P(s)$ 
for an avalanche of
size $s$ presents a series of peaks at 
positions that are proportional to the linear size of the system $L$. 
This is a clear effect of the open boundaries.
Moreover the large peak of order $L^2$ confirms the tendency to
synchronization for $\gamma> 0 $, 
which, however, in this case, the system is not able to sustain
\cite{prl74.118}. 

\begin{figure}[htbp] 
\epsfxsize=2.7truein 
\hskip 0.15truein
\epsffile{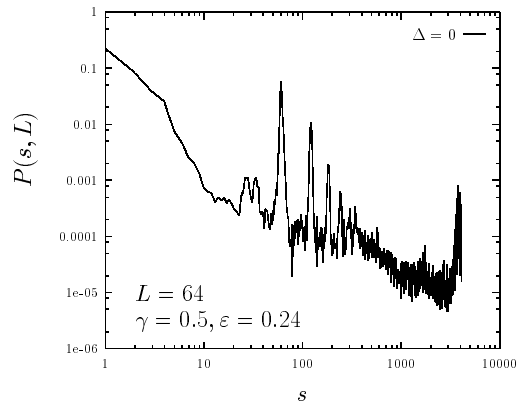} 
\caption{Log-log plot of the stationary distribution 
of avalanche sizes $P(s)$ versus $s$ for the model without diversity.
}
\end{figure}

The situation changes completely for nonidentical oscillators. 
For simplicity we have considered a uniform distribution of periods.
The width $ \Delta$, expressed
as the length of the symmetric interval $(T - \Delta/2, T + \Delta/2)$
centered without loss of generality around $T=1$, 
is a measure of disorder or diversity.
In Fig. 2 we plot the distribution of avalanche sizes 
for different values of $ \Delta$ for the same 
$\gamma$ and $\varepsilon$ as in Fig. 1.
We observe several stages. First of
all, the sequence of peaks displayed in Fig. 1 typical of identical
oscillators continuously disappears when diversity increases.
Then the distribution of avalanches becomes
smoother, without intermediate peaks, but still maintaining the large one
corresponding to avalanches of almost the size of the system ($L^2$),
as displayed for $\Delta=0.15$ in Fig. 2 where
this trend towards synchronization can be seen clearly.
The behavior is 
supercritical, because there are many events able to span the system.
More interesting transitions take place as disorder increases. 
For a larger width, the system self-organizes in a
critical state, without any spatial characteristic scale, as the
power-law distribution of avalanche sizes in the curve 
with $\Delta=0.5$ of Fig. 2 indicates. 
The effect of the different periods is to
reduce the probability of having large avalanches. Then, for very
wide distributions of periods one could expect a strong decay of
$P(s)$. 
In fact, this is what happens when $\Delta \ge 1.5$ for the case of Fig. 2,
as exemplified by the curve with $\Delta = 2$:
a characteristic scale independent of the system size appears
and is responsible of the exponential decay. 
This means that for large diversity avalanches
are localized and the system is subcritical.
These  transitions are not sharp, and the reported values of $\Delta$ 
can change with $\varepsilon$ and $\gamma$. 
In particular the loss of criticality and the appearance of a 
finite correlation length has been found difficult to characterize.
Notice the resemblance 
between the three curves in Fig. 2 and those found in percolation,
where the critical region is restricted to a single point of the control 
parameter \cite{percolation}. However our model shows a finite
region of criticality instead of an infinitesimal one,
as we are going to show below.
This kind of behavior is, for the best of our knowledge,
the first case where 
SOC is found between a supercritical region and a subcritical one
in this class of models.
\begin{figure}[htbp] 
\epsfxsize=2.7truein 
\hskip 0.15truein
\epsffile{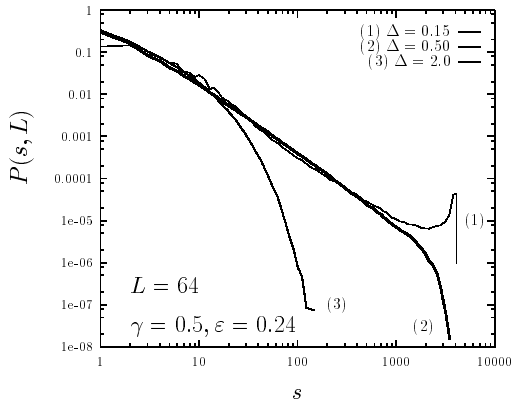} 
\caption{
Same as Fig. 1 for different degrees of diversity. 
}
\end{figure}

Let us pay some attention to the region where the power-law decay of
$P(s)$ is reported. 
We have performed a finite-size scaling analysis for different 
widths $\Delta$. 
The results are shown in Fig. 3. A data collapse for different system sizes
$L$ is obtained when plotting $L^{\beta} P(s,L)$ against the rescaled
variable $s/L^{\nu}$. 
The increment of system size does not show any deviation from the scaling
for separate enough values of the control parameter $\Delta$, 
supporting our statement of a critical region instead of a critical point.
As a complement we plot in Fig. 4 the mean size of the avalanches 
as a function of system size, for different values of diversity.
The behavior $<s> \sim L^{2\nu-\beta}$ 
(consequence of the scaling ansatz) 
even for large $L$ confirms the scaling in the critical region.
In addition, we have released the restriction of 
identical coupling strengths, 
introducing randomness in space ($\varepsilon_{ij}$ quenched random
variable) or in space and time  ($\varepsilon_{ij}(t)$ annealed),
by means of a uniform random distribution around the mean value 
$\varepsilon$.
We have verified that SOC is robust under this perturbation
and hence identical couplings are not a necessary condition
to obtain criticality.
This feature could be relevant in realistic models of spiking neurons.

The results shown so far are 
not characteristic of a particular value of the 
parameters which describe the system. 
In fact there is a region in the $(\gamma,\varepsilon)$-space
where diversity induces SOC and it corresponds to large values of
$\varepsilon$ and small $\gamma$.
It would be very interesting to have knowledge of
the complete phase diagram of the model.
However, taking into account that three parameters are involved: 
$\gamma$, $\varepsilon$, and the width $\Delta $,
and different system sizes are needed,
a complete sweep of the phase space would require an enormous 
effort, that is beyond our possibilities. 
%
Nevertheless, let us mention that for large
$\gamma$ there exists a range of $\varepsilon$ values
which give complete synchronization, in
the sense previously explained, 
no matter the width of the distribution of periods.
%
%
\begin{figure}[htbp] 
\epsfxsize=2.7truein 
\hskip 0.15truein
\epsffile{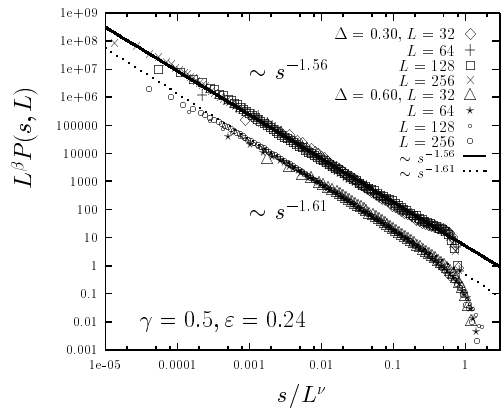} 
\caption{Finite-size scaling analysis of the distribution
of avalanche sizes for the critical region.
For $\Delta=0.3$ we obtain $\nu=2.0$ and $\beta=3.15$ 
(this curve has been shifted one decade upwards
for clarity sake)
whereas for $\Delta=0.6$, $\nu=1.8$ and $\beta=2.85$ are used.
}
\end{figure}

Our results have also sense in the context of earthquakes if we imagine
the Feder and Feder model as a rough version of the Burridge-Knopoff 
spring-block model \cite{BK}.
The different intrinsic periods of each unit will be caused by different
elastic constants in the springs connecting the blocks with the driving plate.
When $\gamma > 0$ a nonlinearity in the elastic response of these
springs is introduced.
With the same goal in mind we have examined our disordered model replacing 
(\ref{ffrule})
by the Olami {\it et al.} (OFC) rules \cite{OFC,prl74.118}, 
and we have found that the SOC region is robust
in spite of a very large disorder, although eventually it can give rise to 
localized avalanches.
These results contrast with the studies performed in Refs. \cite{Jan,Tor}
where a distribution of thresholds was considered.
It was found that while
disorder is irrelevant in the conservative regime, it
destroys criticality for the dissipative case, leading to an exponential
distribution of avalanche sizes \cite{Jan}. 
A similar change in the
collective properties of the disordered system has been used to claim
the lack of robustness of OFC as a model of earthquakes \cite{Tor}. 
Other authors \cite{Ceva} have considered the influence
of defects in the model.
The main result was to observe that SOC is robust
even for a large number of defects. 
Notice that for the same model randomness
included in different parameters leads to different behaviors.

\begin{figure}[htbp] 
\epsfxsize=2.7truein 
\hskip 0.15truein
\epsffile{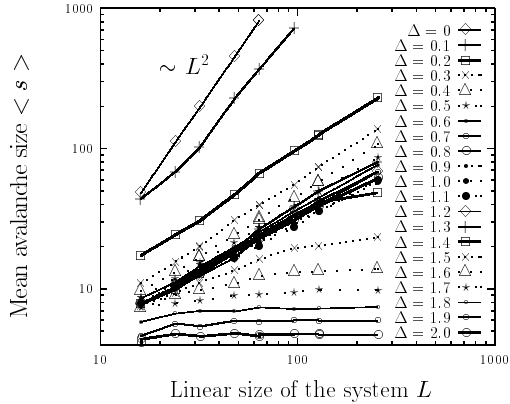} 
\caption{Mean size of the avalanches as a function of the system size
for different values of diversity using $\gamma=0.5$, $\varepsilon=0.24$.
For identical units $<s>$  scales as $L^2$.
This exponent decreases continuously with increasing diversity
in the supercritical region.
Notice the accumulation of data points for a wide range
of  $\Delta$ values (starting for $\Delta \geq 0.3$)
in a narrow interval of mean avalanche sizes, corresponding to
the critical region. 
Here the data fit the scaling $L^{2 \nu - \beta}$.
It is difficult to precise up to which values the scaling holds,
but when $\Delta \geq 1.5$  the growth of $<s>$ with $L$ is 
clearly logarithmic, in agreement with a subcritical region.
}
\end{figure}

Finally, let us
remark on the effect that different
types of noise may have on the collective features of the model.
The original properties of the Feder and Feder model 
(with $\gamma=0$) are not robust to 
noise, e.g., altering the
relaxation rule (\ref{ffrule}) by adding a small random number 
to any reset unit, changes the
cooperative behavior of the system \cite{tesikim,review}. It does not tend
to form a few groups of elements with the same phase, 
but it goes towards a SOC state.
Note that this type of noise has a
completely different nature than the quenched source of 
diversity considered in this paper. 
While the first can be triggered by internal fluctuations,
the second is an inherent feature of each member of the
population. 
Furthermore, while the dynamic noise only induces SOC 
in the linear regime ($\gamma=0$), 
and for very small noise intensities, diversity induces
SOC in a wide region of the parameter space.

In summary, we give an example showing that diversity is a new
mechanism for the emergence of SOC and that criticality in
nonequilibrium systems is not just a singularity in parameter space, 
as it happens in equilibrium. Our results have interest 
for models of integrate-and-fire neurons as well as for earthquakes.

We are indebted with A. Arenas, P. Bak, K. Christensen, 
and B. Tadi\'c for many suggestions and discussions.
A.C. really acknowledges the warm hospitality of Brookhaven National
Laboratory as well as a scholarship of the Spanish MEC.
This work has been supported by CICyT under grant PB94-0891.

Note: Just about submitting this paper we became aware of
Ref. \cite{Mousseau} where disorder
is introduced in the couplings for the OFC model (with $\gamma=0$),
attaining a collective behavior in agreement with our results.

\end{document}